\documentclass[journal]{IEEEtran}
\usepackage{cite}

\ifCLASSINFOpdf
\usepackage[pdftex]{graphicx}
\else
\fi
\usepackage[cmex10]{amsmath}
\usepackage{amssymb}

\newcommand{\gNSVc}%
    {\ensuremath{\raisebox{0.25ex}{\scalebox{1.1}{$\gamma$}}_{\mbox{\footnotesize NSV}}}}

\begin{document}
%
\title{A Conditional Random Field Model for Context Aware Cloud Detection in Sky Images}
%
%
%

\author{Vijai~T.~Jayadevan,
        Jeffrey~J.~Rodriguez,~\IEEEmembership{Senior Member,~IEEE,}
        and~Alexander~D.~Cronin
        }

%
%

\markboth{IEEE Transactions on Geoscience and Remote Sensing,~Vol.~00, No.~0, Month~Year}%
{Jayadevan \MakeLowercase{\textit{et al.}}: A conditional Random Field Model for Context Aware Cloud Detection of Sky Images.}
%



\maketitle

\begin{abstract}
A conditional random field (CRF) model for cloud detection in ground based sky images is presented. We show that very high cloud detection accuracy can be achieved by combining a discriminative classifier and a higher order clique potential in a CRF framework. The image is first divided into homogeneous regions using a mean shift clustering algorithm and then a CRF model is defined over these regions. The various parameters involved are estimated using training data and the inference is performed using Iterated Conditional Modes (ICM) algorithm. We demonstrate how taking spatial context into account can boost the accuracy. We present qualitative and quantitative results to prove the superior performance of this framework in comparison with other methods applied for cloud detection. 
\end{abstract}


%
\IEEEpeerreviewmaketitle

\section{Introduction}
%
%
%
%
\IEEEPARstart{G}{round} based sky imaging (GBSI) systems have become very popular these days for the task of making cloud observations. For instance GBSI systems are employed extensively for predicting intermittency due to clouds in the field of intra-hour solar power forecasting \cite{chow2011,Marquez2013,jayadevan2012}. The Whole Sky Imager (WSI) developed by
the Scripps Institution of Oceanography at the University of California, San Diego \cite{shields1998whole} and the Total Sky Imager (TSI) developed by Yankee Environmental Systems, Inc. \cite{long2006retrieving} are two popular GBSI systems. We \cite{jayadevan2012} like many other research groups around the world \cite{long2006retrieving,pfister2003cloud,cazorla2008development,seiz2007cloud} have developed our own GBSI system. But unlike most other GBSI systems which are stationary and use either an upward looking camera fitted with a fish eye lens or a camera looking down on to a curved mirror to obtain a complete view of the sky, we use a low cost camera fitted with a fish eye lens which tracks the sun. Higher resolution around the sun (which is preferred since we care about predicting solar power intermittency due to clouds) is one of the advantages of this set up. Since it tracks the sun the position of the sun on  the image remains constant and this makes occluding the sun (to prevent saturation of the image) easier. Also the relative area occupied by this occlusion is small in comparison to other imaging systems which is another advantage of this set up.

\begin{figure}[!t]
\centering
\includegraphics[width=3.5in]{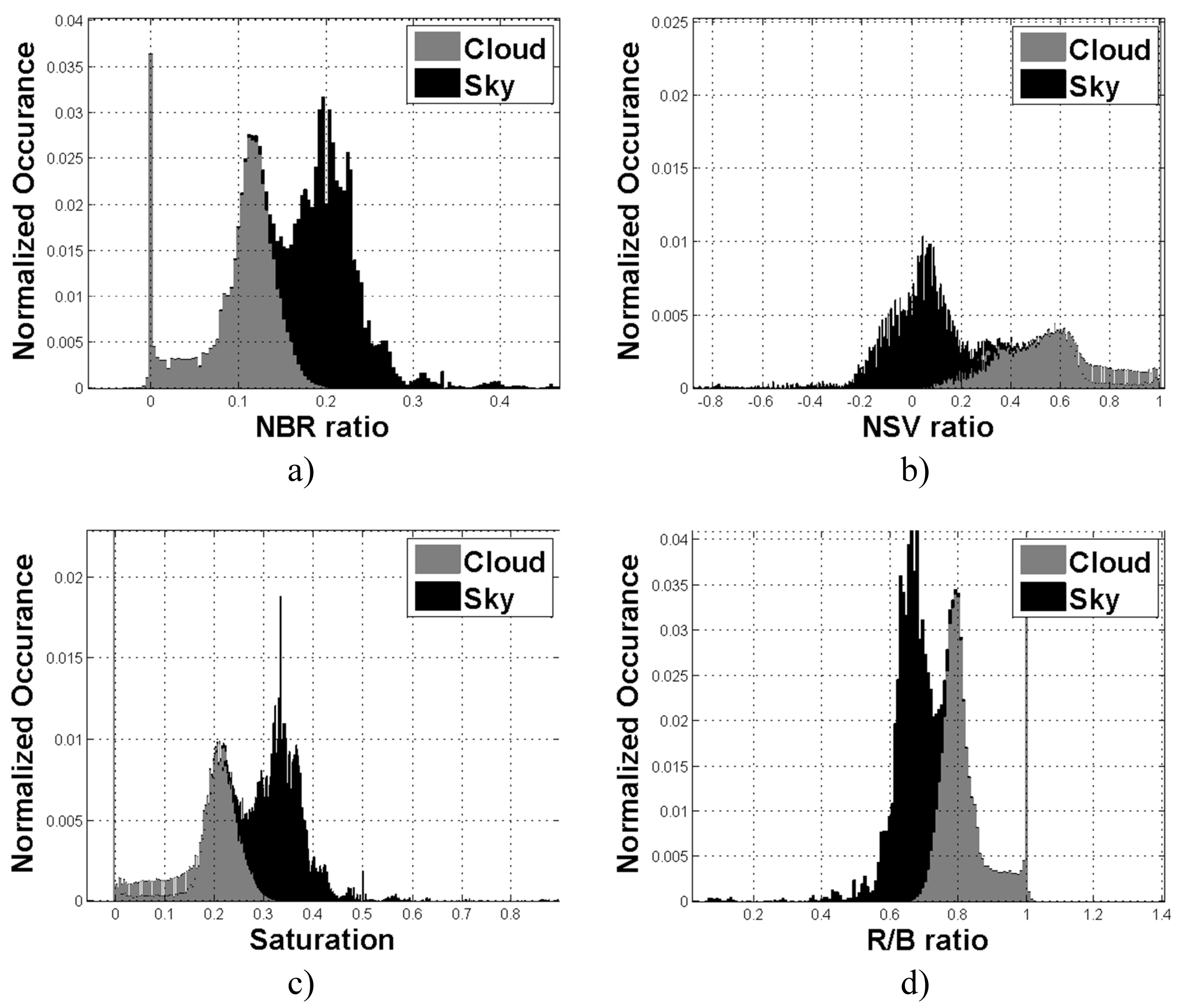}
\caption{Normalized Histograms in various feature spaces.}
\label{fig:Hists}
\end{figure}

Our main goal is to predict intermittency due to clouds for which cloud detection is an important step. One of the earliest works used an empirically derived fixed threshold on the red to blue channel (RB) ratio for cloud detection \cite{johnson1989automated,shields1993automated,shields1998whole}. Another fixed thresholding scheme was suggested by Souza-Echer et al. where they apply a fixed threshold on the saturation component of the IHS colorspace \cite{souza2006simple}. Neto et al.\cite{mantelli2010use} used multidimensional Euclidean geometric distance (EGD) and Bayesian methods to classify sky and cloud patterns based on the observation that sky and cloud patterns occupy different loci on the RGB colorspace. Recently Ghonima et al. \cite{Kleissl2012} proposed the use of the difference in RB ratio between the pixel to be classified and the corresponding pixel in the clear sky library for cloud opacity measurements. They described a method to compensate for the variations in aerosol optical depth by using a haze correction factor derived with the help of a clear sky library. Yamashita et. al \cite{yamashita2004cloud} and Li et al. \cite{li2011} proposed the use of normalized blue to red channel (NBR) ratio (Yamashita et al. called it sky index) for cloud classification and detection. NBR ratio is defined as $${\mathrm{NBR\ ratio} = (B - R)/(B + R)}$$ where ${B}$ and ${R}$ represents the blue and red channel intensities of the pixel respectively. Li et al. also proposed a hybrid algorithm combining fixed and adaptive thresholding schemes. They apply a threshold on the standard deviation of NBR ratio to decide whether to use a fixed thresholding scheme or an adaptive thresholding scheme. If the standard deviation of NBR ratio is below a threshold a fixed threshold determined statistically from training data is used for cloud detection. Otherwise an adaptive thresholding algorithm based on cross entropy minimization is applied to obtain the threshold for the image.\textbf{]]}. In \cite{jayadevan2014NVS} we showed that NSV ratio could serve as a contrast enhancing feature suitable for adaptive thresholding. The NSV ratio is defined as $${\gNSVc = (1-\lambda)/(1+\lambda)}$$ where $${\lambda = S/V}$$ ${V}$ is the value component and ${S}$ is the saturation component in the HSV colorspace. As discussed in \cite{jayadevan2014NVS,li2011} and as depicted in the Fig.~\ref{fig:Hists} there is considerable overlap between cloud and sky pixels in the various feature spaces. Please note that unlike in \cite{jayadevan2014NVS} the histograms shown here corresponds to a set of images used as a training set and not of a single image. These histograms depict the problem associated with a fixed thresholding scheme. Whatever be the threshold that is picked it will inevitably misclassify some pixels. 

On the other hand in the case of adaptive thresholding it is only necessary to ensure that the classes do not overlap in the feature space on an image by image basis. In \cite{jayadevan2014NVS} though we showed that the NSV ratio provides a strong separation between the two classes (cloud and sky) in the feature space for many images, we also mentioned that this is not always true. The dependency of the NSV ratio on the value component of the pixels (in the HSV space) causes the NSV ratio for some dark clouds to overlap with that of sky pixels. We also showed that an adaptive thresholding scheme like cross entropy minimization might pick a wrong threshold if the intra-class variance is high, which is a likely scenario in the case of sky images. Minimum cross entropy thresholding and Otsu's thresholding can be viewed as a surface fitting problem, and the best fit may not always correspond to the best segmentation \cite{jayadevan2014NVS}. 

To overcome these issues, we propose a cloud detection scheme based on the CRF framework. We explain how this model addresses each of the issues mentioned above. Quantitative and qualitative results are presented, and possible extensions to this work are discussed towards the end.

\section{Proposed Method}
Probabilistic graphical models like Markov random fields (MRF) and conditional random fields (CRF) have been extensively applied for the task of contextual image segmentation \cite{Panjwani1995,Held1997,Melgani2003,li2012thin,Shotton2006,Kumar2003,kohli2009}. 
Li et al. \cite{li2012thin} proposed an MRF model for detecting thin clouds. Though MRF models have proven to be successful for computer vision tasks like segmentation and image denoising, they have some drawbacks. Being a generative model, MRF based approaches model the joint density ${P(\textbf{x},\textbf{y})}$, where ${\textbf{x} = \{{x_i}\}_{i \in S}}$ represents the input image data, where ${S}$ is the set of all sites. A site can correspond to a single pixel or a group of pixels (region). And ${\textbf{y} = \{{y_i}\}_{i \in S}}$, ${y_i \in L}$ represents the class labels corresponding to the sites ($\textbf{y}$ is often referred to as a label configuration) and ${L}$ is the set of all possible labels/classes. In the MAP (maximum a posteriori) MRF framework, Bayes' rule is employed to derive the posterior probability of the labels given the data:
$${P(\textbf{y}|\textbf{x}) = \frac{P(\textbf{x},\textbf{y})}{P(\textbf{x})} = \frac{P(\textbf{x}|\textbf{y})P(\textbf{y})}{P(\textbf{x})} \propto P(\textbf{x}|\textbf{y})P(\textbf{y})}$$
In many cases such as in the case of cloud detection ${\textbf{x}}$ is always observed (i.e. during training as well as while performing inference or classifying) and hence it is not necessary to model ${P(\textbf{x})}$ which might not be a simple function. Another drawback of MRF modeling is that often for tractability the likelihood ${}P(\textbf{x}|\textbf{y})$ is assumed to have a factorized form, i.e. ${}P(\textbf{x}|\textbf{y})$ = ${\prod\limits_{i\in S}p(x_i|y_i)}$. This assumption is too restrictive as complex relationships usually exist between data at neighboring sites. And finally the MRF model imposes label consistency (modeled as the prior ${P(\textbf{y})}$) uniformly over the entire image without taking into consideration the observed data.

CRF models, on the other hand, are discriminative models which directly model ${P(\textbf{y}|\textbf{x})}$, the probability of a  label configuration given the data. The CRF model therefore does not attempt to model the distribution of input data ${P(\textbf{x})}$. The conditional independence assumption of the likelihood model given the labels is relaxed in the case of CRFs. CRFs also allow us to model data-dependent label interactions as the clique potentials, or the interaction potentials are functions of both the labels and the data. This property, as we will see, plays a crucial role in cloud detection. Due to the above-mentioned factors, CFRs outperform MRF models for various computer vision tasks including image segmentation \cite{He2004,Kumar2003}.

\begin{figure}[!t]
\centering
\includegraphics[width=3.5in]{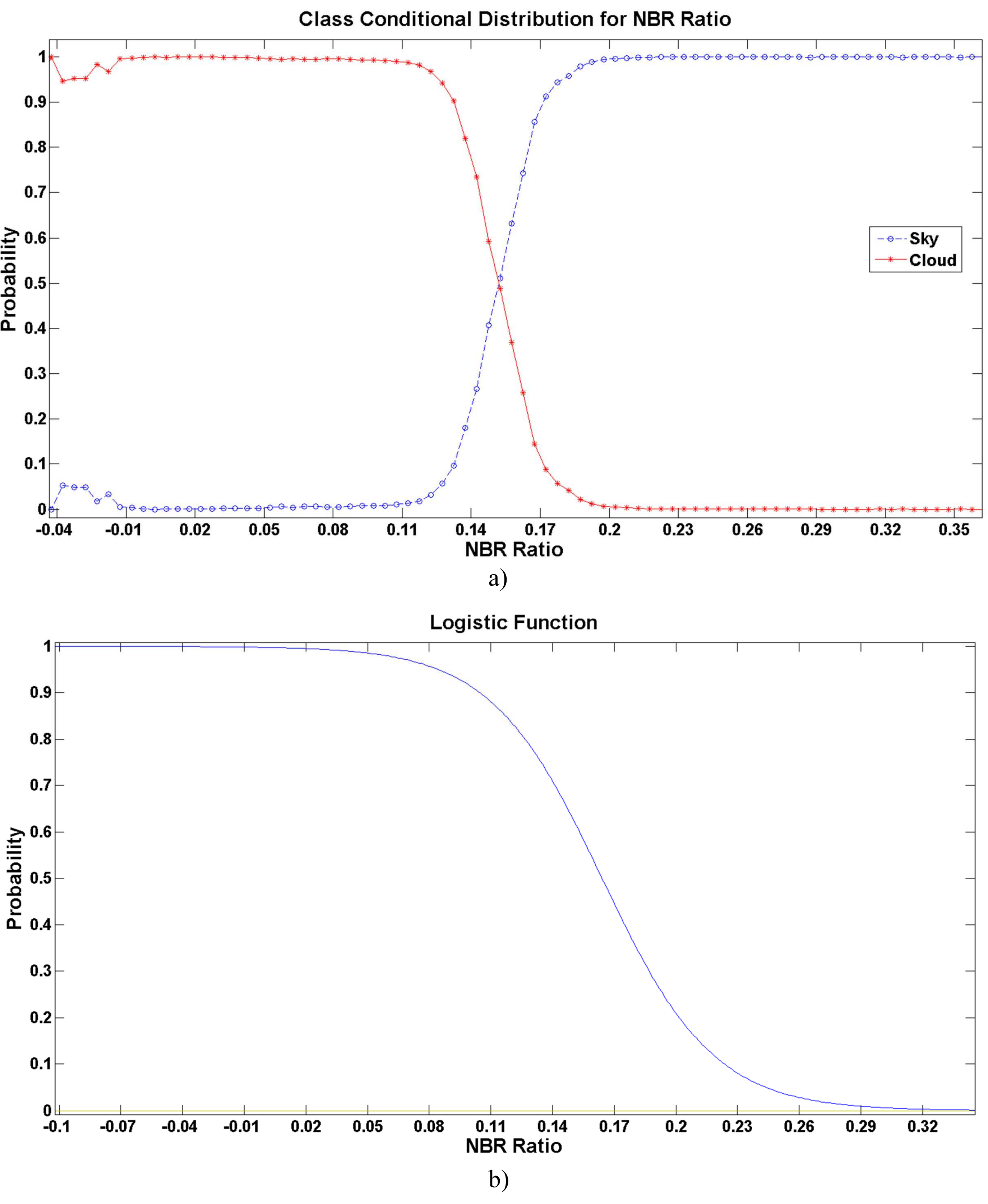}
\caption{a) Class conditional distribution for NBR ratio obtained using training data and b) Logistic function, $\psi$, with parameters ${\alpha_0}$ = 6.072 and ${\alpha_1}$ = -37.001 estimated using training data.}
\label{fig:ccd}
\end{figure}

\subsection{CRF Model}

As mentioned before, we combine a discriminative classifier and a higher order clique potential in a CRF framework. Again ${\textbf{x} = \{{x_i}\}_{i \in S}}$ represents the input image data, where ${S}$ is the set of all sites. We divide the image into homogeneous regions using mean shift clustering \cite{Comaniciu2002}, hence each region defines a site. The labels corresponding to the sites are ${\textbf{y} = \{{y_i}\}_{i \in S}}$, ${y_i \in L}$ with ${L = \{0,1\}}$ where 0 is the label for sky and 1 is the label for cloud. Following the discriminative random field approach \cite{Kumar2003},  assuming $\textbf{y}$ to obey the Markov property conditioned on the data $\textbf{x}$ (i.e., ${P(y_i|\textbf{x}, y_{S - \{i\}}) = P(y_i|\textbf{x}, y_{N_i})}$, where $N_i$ represents the neighborhood of site $i$), and using the Hammersley Clifford theorem \cite{li2009markov} the posterior distribution of the label configuration is defined as
\begin{equation}\label{eq1}
\begin{split}
P(\textbf{y}|\textbf{x})& = \frac{1}{Z}exp \left( \sum_{i\in S} \psi(x_i) + \ \beta \sum_{i\in S}\sum_{j \in N_i} \phi(\textbf{x},y_i,y_j) \right)
\end{split}
\end{equation}
where 
\begin{equation}
\psi(x_i) = \frac{exp(\alpha_0 + \alpha_1K_i)} {1 + exp(\alpha_0 + \alpha_1K_i)}
\end{equation}

\begin{equation}
\phi(\textbf{x},y_i,y_j) = 
\begin{cases}
(V_s - V_i),\ V_s = \overline{V_j}\vert{y_j = 0}\ \text{when}\ y_i= 0\\
(V_i - V_c),\ V_c = \overline{V_j}\vert{y_j = 1}\ \text{when}\ y_i= 1 
\end{cases}
\end{equation}
and
\begin{equation}\label{eq1}
\begin{split}
Z = \sum_{\textbf{y}}exp \left( \sum_{i\in S} \psi(x_i) + \ \beta \sum_{i\in S}\sum_{j \in N_i} \phi(\textbf{x},y_i,y_j) \right)
\end{split}
\end{equation}
${K_i}$ and ${V_i}$ represent the NBR ratio and NSV ratio at site ${i}$, respectively; ${\alpha_0\text{, }\alpha_1\text{ and }\beta}$ are the parameters that need to be estimated. $Z$ is the partition function and serves as a normalization factor. It is computed by summing the model over all possible configurations of $\textbf{y}$.

Both the NBR and NSV ratios are utilized in our CRF model. As can be seen in Fig.~\ref{fig:ccd}, the class conditional density for cloud is almost one for all NBR values below 0.11, and it is almost zero for all values above 0.2. It is mainly in the range [0.11, 0.20] that reliance on NBR alone would result in misclassification, and outside this range we can make cloud detection decisions almost unambiguously. Thus, there are a considerable number of regions which we can classify as cloud or sky with very high confidence. And for the remaining regions where there are ambiguities, we can look at the spatial context to make decisions. As we showed in \cite{jayadevan2014NVS} the NSV ratio exhibits enhanced contrast between cloud and sky, so it could serve as an effective, local feature for testing spatial consistency. But due to the dependency of this ratio on the value component of the HSV space of the image, it makes more sense to use the NBR ratio as a global feature.

\begin{figure}[!t]
\centering
\includegraphics[width=3.5in]{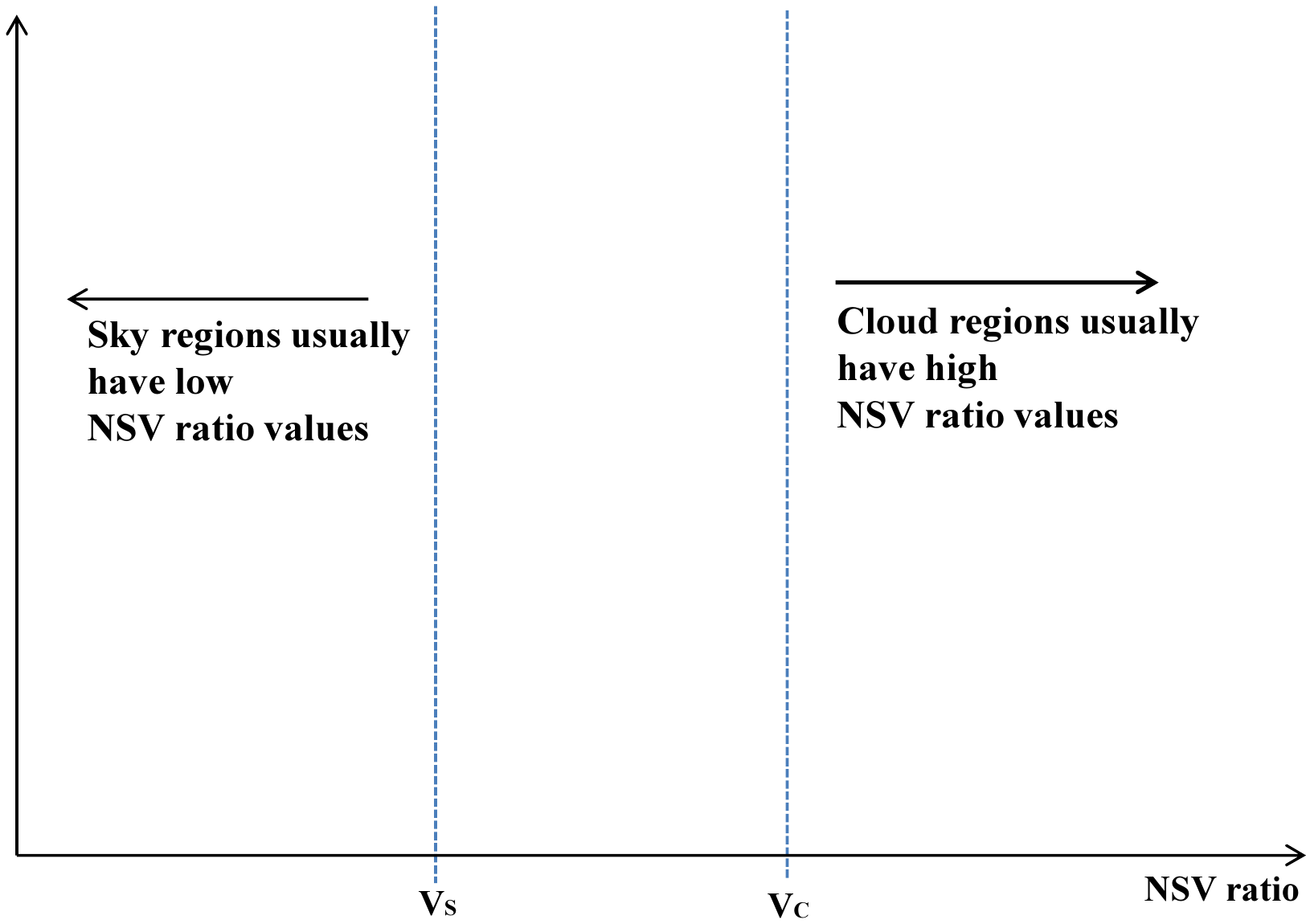}
\caption{Figure illustrating how the interaction potential is defined.}
\label{fig:IneractionPot}
\end{figure}

This idea can be modeled very well using a CRF framework. The function ${\psi}$ is nothing but a logistic regression classifier. This classifier gives a well calibrated probability value indicating whether the site under consideration is cloud or sky just based on the NBR ratio at that site. As shown in Fig. 2, the logistic function derived from the training data is a smoother version of the class conditional density for the cloud. In the DRF framework terminology \cite{Kumar2003} the logistic function is the association potential. Now in order to take the spatial context into account, we define the interaction potential function $\phi$. Before defining the interaction potential we need to define the neighborhood $N_i$ for the site $i$. In our case, we define $N_i$ as all regions which are within a 200-pixel radius of the centroid of site $i$ (the site $i$ is not part of the neighborhood). It is not necessary that a region be entirely within this radius to be considered a neighbor; even if it is partially within this range, it will be considered a neighbor. The interaction potential can now be understood with the help of Fig.~\ref{fig:IneractionPot}. $V_S$ and $V_C$ are the average NSV values of sky neighbors and cloud neighbors, respectively. Clouds usually have higher NSV values in comparison to sky regions, hence $V_C > V_S$. The interaction potential has been defined in such a manner that $V_S$ and $V_C$ serve as reference points. I.e., the farther the NSV ratio of a site is to the left of $V_S$ the higher the probability that it is a sky region, and the farther it is to the right of $V_C$ the higher the probability that it is cloud. If the NSV ratio of a site is in between $V_S$ and $V_C$, then the interaction potential is negative, and the magnitude will depend on its proximity to the two reference points.  

As stated earlier, the mean shift clustering algorithm \cite{Comaniciu2002} is used to form homogeneous regions in an image. The parameters including range bandwidth, color bandwidth, minimum number of pixels in a region, etc., are set manually to produce the best qualitative results. Defining the CRF model over regions instead of pixels has two main advantages. First, it would help in speeding up the inference algorithm (the algorithm used to find the most probable labeling configuration given a new image) and secondly it will help combat noise as we will demonstrate in the results section.

\subsection{Parameter Estimation and Inference}
The parameters in CRF are usually estimated using maximum likelihood estimation (MLE) \cite{elkan2008log,sutton2010introduction,He2004}. MLE finds the parameters that maximize the conditional likelihood of the true labels given the training data. Stochastic gradient descent (SGD) is usually employed to find the MLE estimates \cite{He2004}. In SGD the partial derivatives with respect to each parameter are evaluated. These partial derivatives involve the computation of the expected value of the feature function (for e.g. interaction potential in our case) over all possible label configurations \cite{elkan2008log,sutton2010introduction}. Since this computation is intractable, the expectation is approximated by coming up with an approximation for $P(\textbf{y}|\textbf{x};\theta)$. Here, $\theta$ represents the parameter set \{$\alpha_0,\alpha_1,\beta$\}. Note that $P(\textbf{y}|\textbf{x};\theta)$ is the probability density over all possible label configurations given an image and the parameters, and it needs to be calculated for each image. Loopy belief propagation, Markov chain Monte Carlo (MCMC) and Contrastive divergence are some  methods used for approximate training \cite{elkan2008log,sutton2010introduction}. Being approximations, these methods may not estimate the parameters well \cite{Shotton2006}. Furthermore, for methods like belief propagation, the time complexity is exponential in the size of the largest clique. And in our case, since the CRF model involves higher order cliques, the largest clique might have a size of 80, hence loopy belief propagation cannot be employed. 

An alternative, practical approach is piecewise training \cite{SuttonPiecewise2012,Shotton2006,kohli2009}. In piecewise training each piece of the CRF model is learned independently. In our case, parameters for association potential and interaction potential are learned independently. The training methodology is similar to that of \cite{kohli2009}. Our training data set consists of eight manually labeled images. We first estimate the association potential parameters $\alpha_0$ and $\alpha_1$ using four images from the training set. We use the R project for statistical computing \cite{rcite} to estimate the parameters. The \texttt{glm} function in the stats core package is used for this purpose. This function uses the iteratively reweighted least squares (IWLS) method to obtain the fit. Once the parameters for association potential are estimated, we estimate the parameter for the interaction potential by keeping the association potential parameters constant. The second half of the training dataset (four images) which was not used to train the association potential is used for this purpose. The interaction potential parameter, $\beta$, which minimized the pixel wise classification error (i.e. the total number of misclassified sky pixels and cloud pixels) was picked.

\begin{table}[!t]
\renewcommand{\arraystretch}{1.3}
\caption{Accuracy, Precision and Recall Values with 99.9\% Confidence Interval for Adaptive Thresholding Schemes (ATS) and CRF Based Method }
\label{table_AdVsCRF}
\centering
\begin{tabular}{|c|c|c|c|}
\hline
\bfseries & \bfseries ATS & \bfseries ATS & \bfseries CRF\\
\bfseries & \bfseries (NSV ratio) & \bfseries (NBR ratio) & \bfseries \\
\hline
Accuracy & 0.7979 \ $\pm$\ 0.0963 & 0.7173 \ $\pm$\ 0.1654 & 0.9346 \ $\pm$\ 0.0269\\
\hline
Precision & 0.7782 \ $\pm$\ 0.1552 & 0.7324 \ $\pm$\ 0.1954 & 0.9561 \ $\pm$\ 0.0560\\
\hline
Recall & 0.9047 \ $\pm$\ 0.0838 & 0.9022 \ $\pm$\ 0.1196 & 0.9022 \ $\pm$\ 0.0471\\
\hline
\end{tabular}
\end{table}

\begin{table}[!t]
\renewcommand{\arraystretch}{1.3}
\caption{Accuracy, Precision and Recall Values with 98\% Confidence Interval for Fixed Thresholding Scheme and CRF Based Method }
\label{table_FxVsCRF}
\centering
\begin{tabular}{|c|c|c|}
\hline
\bfseries & \bfseries Fixed Thresholding & \bfseries CRF\\
\hline
Accuracy & 0.8892 \ $\pm$\ 0.0359 & 0.9436 \ $\pm$\ 0.0181 \\
\hline
Precision & 0.9420\ $\pm$\ 0.0378 & 0.9597 \ $\pm$\ 0.0341 \\
\hline
Recall & 0.8236 \ $\pm$\ 0.0540 & 0.9095 \ $\pm$\ 0.0309 \\
\hline
\end{tabular}
\end{table}

Once we have estimated all the parameters, we need an inference algorithm which would find the most probable labeling configuration given an image (data) and the parameters. I.e., we want to find $\textbf{y}^*$ such that
$${\textbf{y}^* = \underset{\textbf{y}}{\operatorname{argmax}}\ P(\textbf{y}|\textbf{x};\theta)}$$
Due to the intractability of exact inference , we use the local search alg, iterative conditional modes (ICM), proposed by Besag \cite{besag1986statistical}. ICM is an iterative procedure that maximizes the local conditional probability. For each site in an image we find
$${y_i = \underset{y_i}{\operatorname{argmax}}\ P(y_i|y_j,\textbf{x})}$$
This procedure is iterated until convergence. ICM requires an initial labeling configuration to begin with. The output of the logistic regression classifier is used as the initial guess. 

During training and inference, there may be situations where a particular site has only one class of sites around it, such that all its neighbors are cloud or all are sky. In such scenarios we use the average sky and cloud NSV values of the entire image instead of the corresponding values of the neighbors. In order to calculate the average sky and cloud NSV values of the entire image, the labeling configuration from the previous ICM iteration is used.

\section{Results}
In order to evaluate the performance of the proposed method, two separate test datasets were used. One of these datasets (Set C) contains 22 images and was used to compare the performance of the CRF-based method with the minimum cross entropy method proposed by Li et al.\cite{li2011}. Li's method uses a hybrid scheme combining fixed thresholding and adaptive thresholding. The decision to use an adaptive threshold or fixed threshold is based on the standard deviation of the NBR ratio. But we observed that this decision rule does not always work well because some clear sky and completely cloudy images have a standard deviation greater than that of some images containing both sky and cloud. Hence, for a fair comparison we have only included images containing both sky and cloud in the first dataset. The second dataset (Set D) contains 26 images in total and was used to compare the performance of the proposed CRF method with that of fixed thresholding proposed by Long et al. \cite{long2006retrieving}. This data set contains all the 22 images in the first dataset with an addition of two completely cloudy and two clear sky images.

\textbf{Replace[[}Accuracy, precision and recall computed based on the confusion matrix are employed as metrics for performance evaluation \cite{kohavi1998glossary}. The metrics are defined below:
\begin{eqnarray*}
\mathrm{ Accuracy} & = & \mathrm{(TP + TN)/(TP + TN + FP + FN)}\\
\mathrm{Precision} & = & \mathrm{TP /(TP + FP)}\\
\mathrm{Recall} & = & \mathrm{TP /(TP + FN)} 
\end{eqnarray*}
Here, TP (true positive) is the number of cloud pixels classified correctly, TN (true negative) is the number of sky pixels classified correctly, FP (false positive) is the number of cloud pixels that got misclassified as sky pixels and FN (false negative) is the number of sky pixels which got misclassified as cloud pixels.\textbf{]]}

\begin{figure}[!t]
\centering
\includegraphics[width=3.5in]{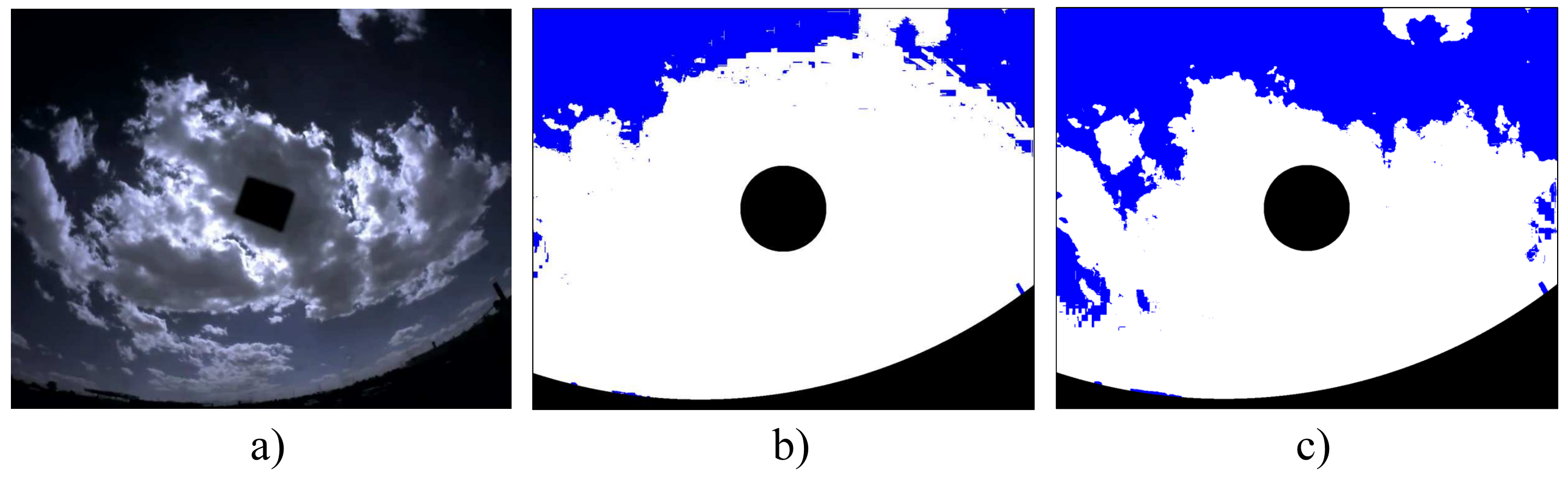}
\caption{a) Original image and segmentation results using adaptive thresholding with b) NBR Ratio and c) NSV Ratio as features.}
\label{fig:ATS}
\end{figure}

\begin{figure}[!t]
\centering
\includegraphics[width=3.5in]{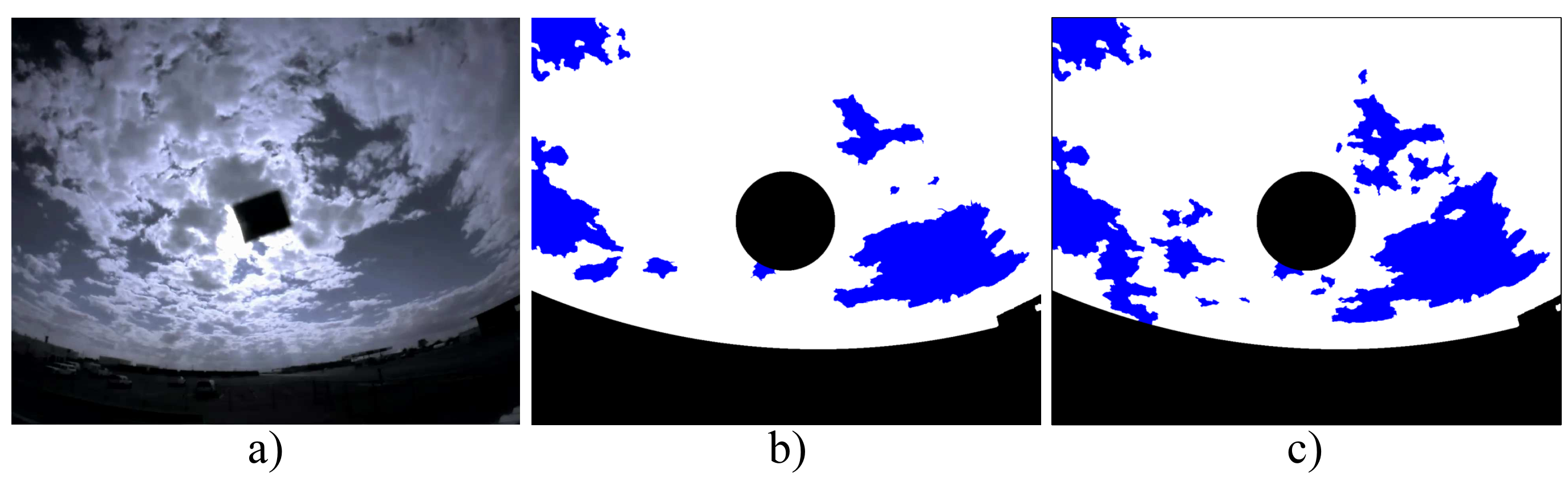}
\caption{a) Original image and segmentation results using the CRF model b) with $\beta$ = 0 and c) with $\beta$ = 0.95 (the estimated value).}
\label{fig:BetaZ}
\end{figure}

\begin{figure}[!t]
\centering
\includegraphics[width=3.5in]{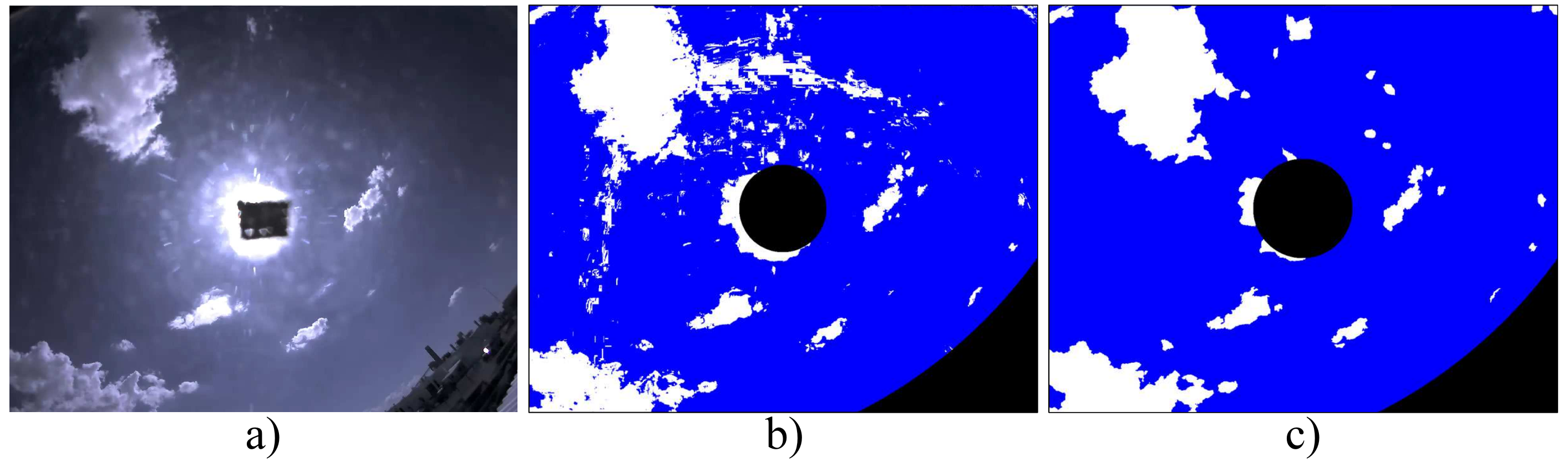}
\caption{a) Original image, b) pixel-wise classification result of logistic regression  and c) segmentation result using the CRF model with $\beta$ = 0.}
\label{fig:MSNoise}
\end{figure}

\begin{figure}[!t]
\centering
\includegraphics[width=3.5in]{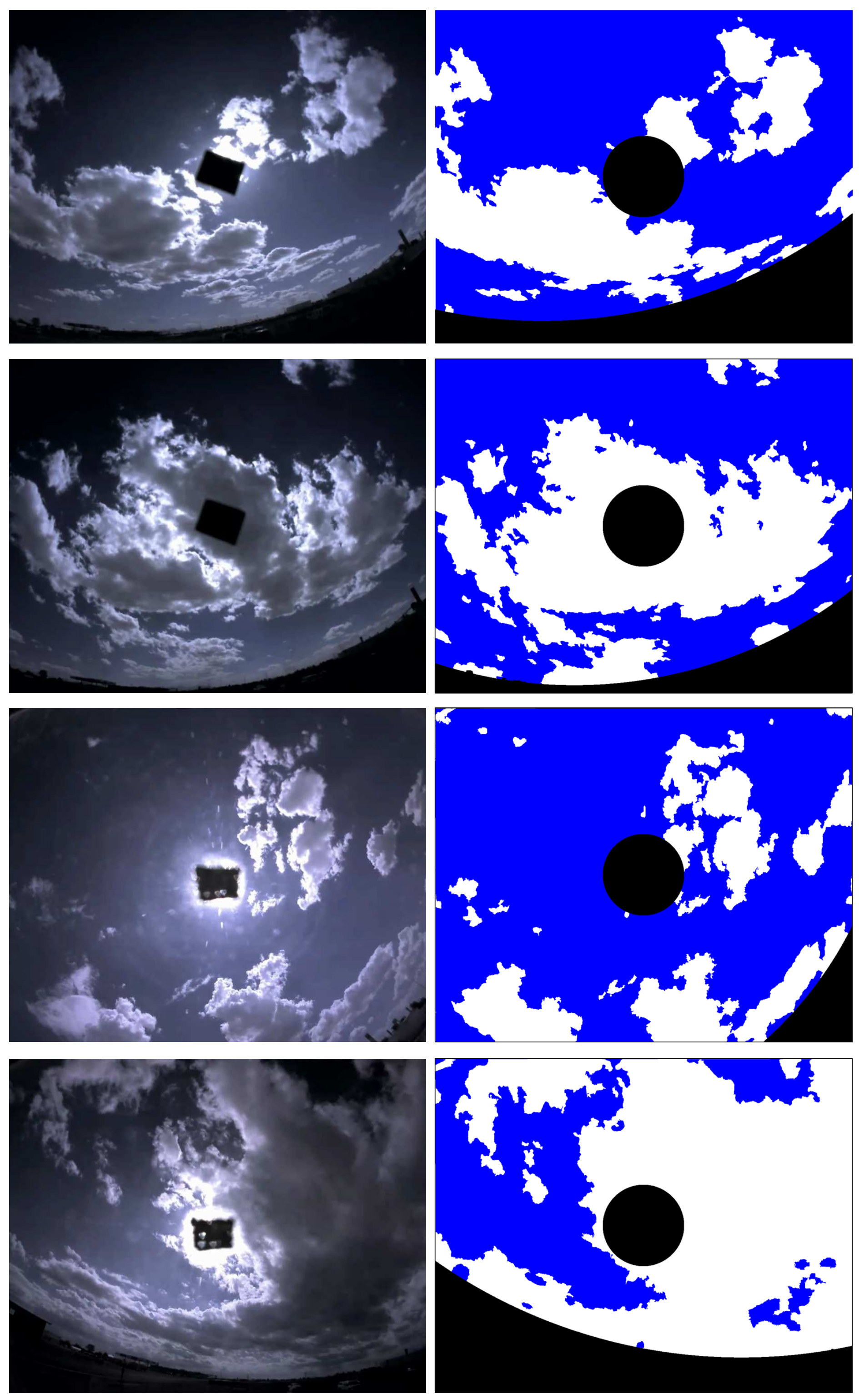}
\caption{Segmentation results obtained using the CRF model for some images from the test database.}
\label{fig:SegResults}
\end{figure}

Table \ref{table_AdVsCRF} shows the results for two adaptive thresholding schemes and CRF method using Set C. Both adaptive thresholding schemes are based on cross entropy minimization, the difference being that one uses NSV as the feature while the other uses NBR. The accuracy of the CRF-based method is much higher in comparison with the other two methods, and the accuracy  values do not overlap even at 99.9\% confidence. The NSV thresholding performed better than NBR thresholding. The reason for the poor performance of adaptive thresholding is that the best fit does not always correspond to the best segmentation/classification , especially when there is high intra-class variance. For instance, as shown in Fig.~\ref{fig:ATS}, the sky region at the top portion of the image is dark in comparison to the sky region at the bottom of the image. And when the adaptive thresholding algorithm tries to fit a binary surface that minimizes the cross entropy, it ends up misclassifying the entire bottom portion of the image.

The results of comparing the performance of our method with the fixed thresholding scheme is provided in Table \ref{table_FxVsCRF}. Though the fixed thresholding scheme does better than adaptive thresholding, the CRF method provides considerable improvement in accuracy. The accuracy values do not overlap at 98\% confidence level, so we can be 98\% confident that  the CRF method is better. Here, taking the spatial context into consideration leads to better accuracy. Fig.~\ref{fig:BetaZ} shows that the inclusion of interaction potential helps in correcting some mistakes made in segmentation using  the association potential alone. Looking at the local context is really helpful in regions where the histograms of sky and cloud regions overlap in the feature space. 

Forming regions using mean shift segmentation and then performing region-based classification helps in dealing with noise. Fig.~\ref{fig:MSNoise} shows that using logistic regression (association potential) to classify the image on a pixel-by-pixel basis yields noisier results than classifying regions formed by mean shift segmentation. Hence, defining the CRF model over regions not only speeds up computation but also helps in combating noise. However, in cases where ta cloud has an unusual shape, segmentation may fail to find the proper regions. But for most cases, a region based approach works well. Fig.~\ref{fig:SegResults} shows the segmentation results for some images obtained using our CRF method.

\section{Conclusion}
We have presented a CRF model for cloud detection on ground-based sky images which  outperforms Li's adaptive thresholding algorithm and Long's fixed thresholding algorithm on our images. Though our model uses only two features (NSV ratio and NBR ratio) it is very flexible, and more features (if found useful) can be added very easily. For instance, adding more features into the association potential is trivial. Similarly, the interaction potential could be replaced by one or more features that may help take the local context into account in a better way. Since each imaging system is different, it is possible  to find features different from what we have used which are better suited for different imaging setups. But our aim here is to demonstrate that the idea of combining a discriminative classifier with a higher-order clique potential in a CRF framework is a really powerful scheme for cloud detection.


%

\section*{Acknowledgment}
We would like to acknowledge Tucson Electric Power (TEP), The Arizona Research Institute for Solar Energy Research (AZRISE) and Arizona TRIF program for supporting this work.

\ifCLASSOPTIONcaptionsoff
  \newpage
\fi



%
%
%
\bibliographystyle{IEEEtran}
\bibliography{IEEEabrv,CRFbib}
%

%
%
%




\end{document}